\def\aj{{AJ}}

\def\apj{{ApJ}}

\def\apjs{{ApJS}}

\def\mnras{{MNRAS}}
\def\nat{{Nature}}

\def\pasj{{PASJ}}
\def\aap{{A\&A}}
\def\araa{{ARA\&A}}

\newcommand{\bea}{\begin{eqnarray} }
\newcommand{\eea}{\end{eqnarray}}
\newcommand{\mpc}{$M_\odot\, \rm{pc}^{-2}$}

\documentclass[cup5b]{caps}
\usepackage{graphicx}
\usepackage{amssymb}
\usepackage{ociwsymp1}
\HeadText{K. Wada}

\def\plotone#1{\centering \leavevmode
\includegraphics[width=.95\columnwidth]{#1}}

\def\farcs{\hbox{$.\!\!^{\prime\prime}$}}

\begin{document}

\pagenumbering{arabic}

\author[]{KEIICHI WADA\\National Astronomical Observatory of Japan, Tokyo}

\chapter{Fueling Gas to the Central Region \\ of Galaxies}

\begin{abstract}
Supplying gas to the galactic central regions is one of key ingredients for 
AGN activity. I will review various fueling mechanisms for a $R\approx 0.1$ kpc 
region, determined mainly by numerical simulations over the last decade.  I 
will also comment on the bars-within-bars mechanism.  Observations suggest 
that the stellar bar is not a sufficient condition for gas fueling.  Moreover, 
considering the various factors for the onset of gas accretion, stellar bars 
would not even be a necessary condition. I introduce recent progress obtained 
through our two- and three-dimensional, high resolution hydrodynamical 
simulations of the ISM in the central 0.1--1 kpc region of galaxies.
Possible structure of the obscuring molecular tori around AGNs is also shown.
The nuclear starburst is an important factor in determining the structure of
the molecular tori and the mass accretion rate to the nucleus.  It is natural 
that the ISM in the central 100 pc region is a highly inhomogeneous and 
turbulent structure. As a result, gas accretion to the central parsec region 
should be time dependent and stochastic.  The conventional picture of gas 
fueling and the AGN unified model may be modified in many respects.

\end{abstract}

\section{Conventional Picture of the Fueling Problem}
\label{wada_sec: 1}
Accretion of gas to the supermassive black hole in the galactic center is 
the source of all AGN activity. A long-standing issue concering this gas 
supply is the ``fueling problem''  --- that is, the question of how to remove 
the large angular momentum of the gas in a galactic disk and
funnel it into the accretion disk in the central AU region.  This was one 
of the main topics at the ``Mass-transfer Induced Activity in Galaxies'' 
conference held in Lexington in 1993 (Shlosman 1994).  The well known cartoon 
by E. S. Phinney of a baby being fed by a huge spoon, published in the 
workshop's proceedings, well represents the essence of the fueling problem.  
To power an AGN luminosity of $\sim 10^{10-11}\, L_\odot$, we need a mass 
accretion rate of $\sim 0.1\, M_\odot$ yr$^{-1}$ with a $\sim 10$\% energy 
conversion rate.  Therefore, to maintain the AGN activity during its lifetime 
of $10^8$ yr, a large amount of the gas, $10^7\, M_\odot = 0.1\, M_\odot$ 
yr$^{-1} \times 10^8$ yr, must be funneled into the black hole. The galactic 
disk is probably a
reservoir of the gas, and since the time scale of $10^8$ yr is comparable to 
the rotational time scale of galaxies, it is natural to postulate that 
the gas is accumulated from the galactic disk.
A number of mechanisms for removing the angular momentum of the gas 
have been proposed. Among them, the use of gravitational torques due to 
galaxy-galaxy interactions (e.g., major/minor mergers and close encounters) or
stellar bars have been considered reasonable means 
of removing angular momentum of the rotating gas.

This is the {\it conventional} picture of the fueling problem. 
After the comprehensive review paper on this subject 
by Shlosman, Frank, \& Begelman (1990), there were many findings, mainly through
numerical simulations. In the next section, I will review various fueling
mechanisms on a scale from $R \approx 1$ kpc down to 100 pc, found in the last decade.
I will show a revision of the ``fueling flowchart'' (Shlosman et al. 1990)
for cases with and without inner Lindblad resonances.
The ``bars-within-bars'' hypothesis, 
which has been considered as an important mechanism to connect
the large scale and small scale, is discussed in \S~\ref{wada_sec: 2.3}.
In \S~\ref{wada_sec: 3},  I will introduce our recent work
on the dynamics and structure of the gas in
the central 100 pc around a supermassive black hole. 
Finally, I will summarize a new picture for gas fueling and discuss
implications from recent observations in \S~\ref{wada_sec: 4}.

\section{The Fueling Flowchart and Its Revision}
\label{wada_sec: 2}
\subsection{Fueling Processes with ILRs}
\label{wada_sec: 2.1}

Shlosman et al. (1990) proposed a ``flowchart'' that describes possible 
fueling mechanisms from 10 kpc down to the central black hole scale.  After 
the review was published, there was a great deal of theoretical and numerical 
work on the gas dynamics in a bar potential, especially on the scale from 1 
kpc to 100 pc. Shlosman et al.'s fueling flowchart shows that if there are 
inner Lindblad resonances (ILR), a starburst ring is triggered at 
$R \approx 1 $ kpc, a ringlike, dense region of gas formed due to the 
resonance-driven mass transfer.  However, the gaseous response to the 
resonances and the final structure of the gas depend not only on the existence 
of Lindblad resonances, but also on the {\it type} of Lindblad resonances. 
Three types of resonance-driven fueling processes were proposed.

(1) If a galaxy has a rigidly rotating central region, two Lindblad resonances 
are expected to exist around the core radius, depending on the pattern speed 
of the non-axisymmetric gravitational potential, $\Omega_p$.  Owing to the two 
ILRs, the inner ILR (IILR) and the outer ILR (OILR), an oval gas ring is 
formed near the two resonances, and if the ring is massive enough, the ring 
fragments due to gravitational instability. This would cause a ringlike 
starburst. Wada \& Habe (1992), however, showed that if the gas mass is 
greater than about 10\% of the dynamical mass, the fragmented oval gas ring 
finally collapses.
As a result, a large amount of the gas is supplied to the central 100 pc.
In this process, the stellar bar removes the angular momentum of the gas,
and self-gravity of the gas also plays an essential role.
Energy dissipation due to shocks caused by collision of the clumps is also a 
key physical process in changing the oval gas orbits
into inner circular motion (see also Fukunaga \& Tosa 1991).
Interestingly, similarly elongated gas rings are found in more complicated
situations, for instance in numerical simulations of stellar bars with a 
gas component and galaxy-galaxy encounter systems (Friedli \& Benz 1993; 
Barnes \& Hernquist 1996).

(2) Related to the IILR, another fueling mechanism for $R\approx 100$ pc is 
possible.  If the gas disk inside the IILR is massive, there is a more rapid 
fueling process, by which the gas can fall toward the center well before the 
ILR ring is formed (Wada \& Habe 1995). The gaseous oval orbits near the ILRs 
are oriented by 45$^\circ$ with respect to the bar potential,
and the distortion of the orbits strengthens with time due to the 
self-gravity of the gas and loss of angular momentum. Eventually, radial
shocks are generated along the major axis of the elliptical orbits.
When the gas on the elongated orbits rush into the shocked region, their orbits
drastically change in the radial direction, and the gas falls toward the 
center.  The gas in the shocked region effectively loses its angular momentum 
because the torque distribution exerted by the bar potential is negative in
the first and third quadrants (assuming that the bar major axis is located 
along the $x$-axis, the gas rotates counter-clockwise), and also because the 
torque is maximum at an angle of $45^\circ$ to the bar major axis (see Fig. 13 
in Wada \& Habe 1995). 

(3) The third type of bar-driven fueling is related to the so-called nuclear 
ILR (nILR).  The nILR appears when there is a central mass concentration, such 
as a super massive black hole or a stellar/gaseous core, in a weak bar 
potential.  For such a mass distribution, the linear resonance curve 
monotonically declines with radius in the region where the central mass 
dominates, such that $\Omega (R) - \kappa (R)/2 = \Omega/2 \propto R^{-3/2}$, 
where $\kappa$ is the epicycle frequency. Using two-dimensional, 
non-self-gravitating SPH (smoothed particle hydrodynamics) simulations, 
Fukuda, Wada, \& Habe (1998) showed that if there is a nILR, offset shocks 
(ridges) are formed around the nILR. The offset shocks extend to the inner 
region and connect to a nuclear ring or to a pair of spirals.  The offset 
spirals and ringlike structure of the molecular gas are often observed in the 
nuclear region of spiral galaxies, for example in IC 342 (Ishizuki et al. 
1990) and NGC 4303 (Schinnerer et al. 2002). The famous ``twin-peak'' 
structure of CO found in M101, NGC 3351, and NGC 6951 (Kenney et al. 1992) 
probably corresponds to the inner region of the resonance-driven offset 
ridges and the ring. 

The gas  loses its angular momentum as well as energy at the trailing shocks, 
which are observed as ridges in the molecular gas or dust lanes.  After 
passing through the shocks several times, the gas settles in the nuclear ring, 
where the orbital energy is in a minimum state for a given angular momentum.
The location of the ring is typically a few times smaller than the radius of
the nILR, but no further mass inflow beyond the ring is expected, provided 
that the self-gravity of the gas is not important. Fukuda, Habe, \& Wada 
(2000) showed that the nuclear spirals and ring can be unstable to 
gravitational instability; the gas ring fragments to many dense clumps, and 
eventually the clumps fall to the center.  Finally, a nuclear dense core, 
whose size is typically about 50 pc, is formed.

One should note that the offset shocks and ring can be also formed around the 
OILR, as originally found by Sanders \& Tubbes (1980) and van Alvada (1985) 
(see also Athanassoula 1992; Piner, Stone, \& Teuben 1995; Maciejewski et al. 
2002). This is because its dynamical character is the same as the nILR. The 
phase shift of oval orbits near the nILR has a radial dependence similar to 
that for the OILR; therefore, trailing spirals are formed around both 
resonances (Wada 1994).  This means that the observed offset ridges and rings
could be the resonant structure formed around the outer ILR. This is the case 
when there is a stellar cusp in the galactic center, which is in fact observed 
in many late-type spiral galaxies (Seigar et al. 2002;
Carollo 2003).  Recent surveys of molecular gas in
the central region of spiral galaxies also suggest that 
the rotation curves in the central kpc region
often show a steep rise (Sofue et al. 1999; Sofue \& Rubin 2001).

Considering these facts, it would be natural that the resonant structure 
driven by the IILR, leading spirals, is not observed, because there is no IILR 
without a central, rigidly rotating region\footnote{Another reason why the 
leading spirals are not observed is that such spirals are dynamically 
unstable, and they evolve into an oval ring in a rotational time scale.}. 
See also Yuan, Lin, \& Chen (2003) on the sensitivity of spiral patterns 
to rotation curves.

\subsection{Fueling Processes without ILRs}	
\label{wada_sec: 2.2}
As shown in the fueling flowchart by Shlosman et al. (1990), the mass of the 
gas disk, or its self-gravity, is a key ingredient in determining the fate of 
gas disks without ILRs.  Shlosman, Frank, \& Begelman (1989) and Shlosman et 
al. (1990) suggested the importance of the bar-mode instability of the gas 
core accumulated by a large-scale stellar bar.  The criterion for the bar-mode 
instability of a rotating disk is $T_{\rm rot}/|W| > t_{\rm crit} $, where 
$T_{\rm rot}$ is the kinetic energy of rotation, and $W$ is the gravitational 
energy. I will discuss this bars-within-bars mechanism in 
\S~\ref{wada_sec: 2.3}.

Even if the gas disk/core is stable against the global bar-mode instability, 
the gas disk could be unstable on a local scale, because the radiative cooling 
is very effective in massive gas disks; thus, the disk can fragment into many 
clumps on the scale of the Jeans length. The clumps suffer from dynamical 
friction with the stellar component, and they eventually fall toward the 
center. Such a process was revealed by three-dimensional $N$-body and 
sticky-particle simulations by Shlosman \& Noguchi (1993) and Heller \& 
Shlosman (1994).  They found that when the gas mass fraction is less than 
about 10\%, the gas is channeled toward the galactic center by a growing 
stellar bar.  For higher gas fractions, the gas becomes clumpy.
Dynamical friction between gas clumps and stars also contributes to
heating of the stellar system. As a result, the growth of the stellar bar is damped.

More recently, high-resolution, hydrodynamical simulations of a massive disk, 
taking into account self-gravity and radiative cooling below 100 K, 
revealed that even if the gas disk is globally stable 
it can be  highly inhomogeneous and turbulent on a local scale
(Wada \& Norman 1999, 2001; Wada, Meurer, \& Norman 2002).
Wada \& Norman (1999, 2001) presented a high-resolution numerical
model of the multi-phase ISM in the central 2 kpc region of a disk
galaxy with and without energy feedback from massive stars.
Using $2048^2 - 4096^2$ grid cells, they found that a globally
stable, multi-phase ISM is formed as a natural consequence of the
non-linear evolution of thermal and gravitational instabilities in the
gas disk.  The surface density ranges over 7 orders of magnitude, from
$10^{-1}$ to $10^{6}$ \mpc, and the temperature extends over 5 decades, from 
10 to $10^6$ K.  They also find that, in spite of its very complicated spatial 
structure, the multi-phase ISM exhibits a one-point probability density 
function that is a perfect log-normal distribution over 4 decades in density.
The log-normal probability density function is very robust even in regions 
with frequent bursts of supernovae. The radial profile of the turbulent disk 
changes to a steeper one in a time scale of $\sim 10^8$ yr (Fig. 11 in Wada \& 
Norman 2001).  The mass inflow is caused by turbulent viscosity 
(e.g., Lynden-Bell \& Pringle 1974).

The turbulent nature of the self-gravitating gas disk was studied in detail by
Wada et al. (2002). They 
found that the velocity field of the disk in the non-linear
phase shows a steady power-law energy spectrum over
3 orders of magnitude in wave number.
This implies that the random velocity field can be modeled
as fully developed, stationary turbulence.  Gravitational and thermal
instabilities under the influence of galactic rotation
contribute to form the turbulent velocity field.
The effective Toomre $Q$ parameter, in the non-linear phase, exhibits a wide
range of values, and gravitationally stable and unstable regions are distributed
in a patchy manner in the disk.
These results suggest that large-scale
galactic rotation coupled with the self-gravity of the gas
can be the ultimate energy source that maintains the turbulence in the
local ISM. Therefore, mass inflow is naturally expected in a 
rotating gas disk. We just need self-gravity of the dense gas and
radiative cooling for fueling. 

\begin{figure}
\centering
\includegraphics[width = 10cm]{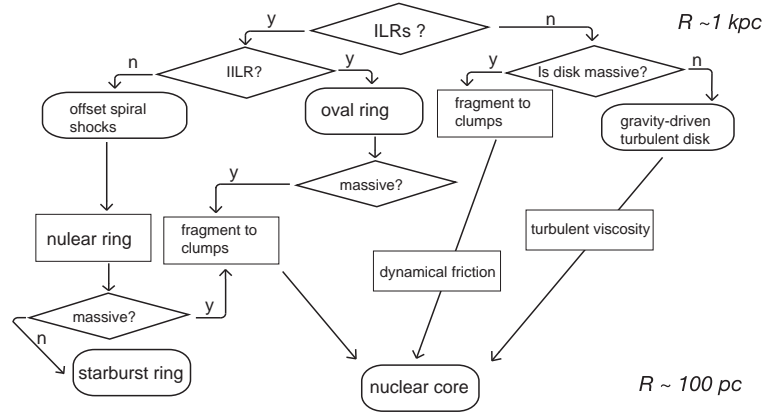}
\caption{A flowchart of various fueling mechanisms from $R\approx 1$ kpc to
100 pc, which is a partial revision from Shlosman et al. (1990), considering
numerical results in the last decade. Shlosman et al. proposed the
``bars-within-bars mechanism'' to connect the scales from $R \approx 100$ pc
to 10 pc.  }
\label{wada_fig: diagram}
\end{figure}

\subsection{A Revised Fueling Flowchart}
In summary, the fueling flowchart from $R \approx 1$ kpc to 100 pc can be 
revised as as shown in Figure \ref{wada_fig: diagram}\footnote{Note that this 
flowchart is not complete.  For example, Maciejewski et al. (2002) proposed a 
fueling mechanism due to nuclear spiral shocks in a non-self-gravitating gas 
disk with a high sound speed. There should be many other factors that determine 
the fueling processes. One should also realize that the diagram is rather 
qualitative. The condition ``massive or not'' depends on the temperature and 
velocity dispersion of the gas, and it is not a simple function of the gas 
mass fraction relative to the dynamical mass. All paths are theoretically 
possible, but it does not say which path is more probable in real galaxies. 
The time scale for each path is also different.  Another point one should keep 
in mind is that gaseous response to the  resonances is not ``discrete.'' 
Because of the dissipational nature of the gas, resonant structures, such as 
spirals, can be formed even if the linear resonance condition (e.g., 
$\Omega_p = \Omega - \kappa/2$) is {\it not}\ strictly satisfied (Wada 1994).}.
This shows that there are many possible fueling paths for the 
$R \approx 100$ pc scale. Dynamical resonances are the key in this diagram, 
since the redistribution of the gas components seems to be sensitive to them.
However, some remarks are necessary.  The effects of resonances on the gaseous 
orbits cannot be ignored, even if there are {\it no} ILRs, namely when 
$\Omega_p > {\rm max}(\Omega - \kappa/2)$.  In this sense, the linear 
condition of the resonances is not a strict criterion for the response of the 
gas to the bar potential (see details in Wada 1994).  Another point to 
keep in mind is that the $\Omega -\kappa/2$ diagrams determined from 
observations can have large errors.  Both $\Omega$ and $\kappa$ are determined 
from rotation curves, but the rotational velocity, especially in the central 
region of galaxies, is not determined accurately, because of the low spatial 
resolution and non-circular motion of the gas.  One should be especially 
careful when rotation curves are obtained from position-velocity diagrams of 
molecular gas in galaxies (Takamiya \& Sofue 2002; Koda \& Wada 2002). 
Finally, gas dynamics in a live stellar bar, where there is a back reaction 
from the gas to the stellar system, could be more complex than that expected 
from gas dynamics in a fixed bar potential (e.g., Heller \& Shlosman 1994). 
 
The mechanisms described in Figure \ref{wada_fig: diagram} are not the 
only ones pertaining to angular momentum transfer. Other mechanisms are 
possible, for example spiral density waves (Goldreich \& Lynden-Bell 1965a,b; 
Lynden-Bell \& Kalnajis 1972; Emsellem 2003), galactic shocks (Fujimoto 1968; 
Roberts 1969), minor mergers (Hernquist \& Mihos 1995; Taniguchi \& Wada 1996),
and the rotational-magnetic instability (Sellwood \& Balbus 1999). 

By means of one or a combination of these mechanisms, 
dense gas cores at $R \approx 100$ pc in the galactic center can be
formed on a dynamical time scale ($\sim 10^{7-8}$ yr). 
This seems to be consistent with the fact that many spiral galaxies are
molecular gas rich in the central region (Sakamoto et al. 1999).
Therefore, a real issue for the fueling problem and AGN activity
is {\it inside} 100 pc.  Recall the original fueling flowchart. 
The only channel toward the galactic center is the ``bars-within-bars''
mechanism, which was proposed by Shlosman et al. (1989).
In the next section, I will review this idea briefly.

\subsection{Remarks on the Bars-within-bars Mechanism}
\label{wada_sec: 2.3}

Shlosman et al. (1989) proposed a novel idea, the bars-within-bars 
mechanism, for fueling AGNs. They used the criterion for bar instability, 
namely that when the ratio between rotational energy and gravitational energy, 
$t_{\rm crit} \equiv T_{\rm rot}/|W|$, is larger than some critical value, 
typically 0.14 for an $N$-body system or 0.26 for an incompressible gas 
sphere.  They rewrote the criterion as $a_{\rm star}/a_{\rm gas} > 
C(t_{\rm crit})/g^2$, where $a_{\rm star}$ is the scale length of the stellar 
system, $a_{\rm gas}$ is the core radius of the gas,  $C(t_{\rm crit})$ is a 
function of $t_{\rm crit}$, and $g$ is the gas mass fraction relative to the 
total mass.  Using this criterion, Shlosman et al. answered the question:
Given $g$ and $t_{\rm crit}$, how much does the gas disk shrink radially to 
become unstable? For example, for $g$= 0.2 and  $t_{\rm crit}$ = 0.14, the 
disk becomes bar unstable when the gas disk shrinks to 1/10 of the 
size of the stellar core.  This means that if the large-scale stellar bar 
sweeps the gas inward to about 1/10 of the bar size ($\sim$ a core radius of 
the stellar disk), then the accumulated gas disk becomes bar unstable.
They claimed that the resulting inflow can extend all the way into
the inner $\sim$10 pc. 

Their argument on the bar instability is reasonable, but they did not actually 
provide a quantitative discussion on the ``resulting inflow.'' Once the core 
becomes bar unstable, they expect that some fraction of the gas will falls 
toward the central region, where viscosity-driven flow would dominate further 
inflow, i.e. toward $R \approx 10$ pc.  However, this has not been 
theoretically or numerically proven.  The redistribution of the mass due to 
the bar instability can drive only a very small part of the gas into $1/10$ of 
the initial radius.  For example, the hydrodynamical simulations of rotating 
gas spheres by Smith, Houser, \& Centrella (1996) show a process of recurrent 
bar instabilities.  As a result of the first bar instability, redistribution 
of the gas takes place; outer spirals are formed, which transfer angular 
momentum outward. The second and subsequent instabilities are much weaker than 
the first one; therefore, the resultant mass distribution is not very 
different from the initial one.  This is because, in order to complete the 
loop, a large part of the angular momentum must be transferred outward with a 
small fraction of the mass; otherwise the fraction  $g$ becomes too small.
Therefore, the gas disk must shrink to a very small radius to satisfy the
instability criterion again.

A serious problem here is how to remove a large part of the angular momentum.
This would be impossible without external mechanisms of transferring the 
angular momentum, such as a secondary bar.  In this sense, the 
bars-within-bars mechanism does not solve the angular momentum 
problem.\footnote{The term ``bars-within-bars'' is used differently in the 
literature.  In the original Shlosman et al. (1989)  paper
the term actually means ``gas bars-within-stellar bars.'' Recently 
many authors use the term to describe ``stellar bars-within-stellar bars.''
See also a recent review paper by Maciejewski (2003a) on this subject.}
A nested stellar bar on a small scale could work to remove angular momentum 
of the bar-destabilized gas. However, as mentioned in \S~\ref{wada_sec: 2.2},
a stellar system can be dynamically heated up due to the interaction between 
the gas components and the stars, and stellar bars in the nuclear region,
where it is especially affected by the dissipation of the gas, can be dissolved
when the gas mass fraction is large enough.

\begin{figure}
\centering
\plotone{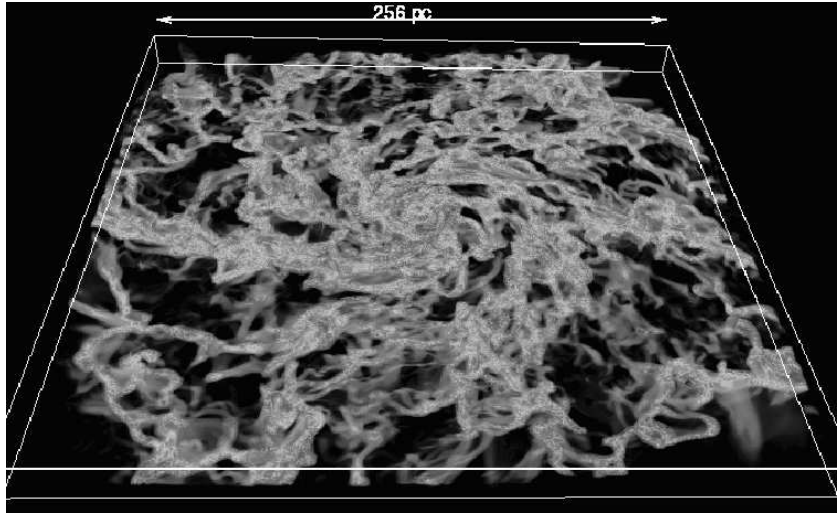}
\caption{Three-dimensional density structure of the ISM in the
central region of a galaxy (Wada 2001).}
\label{wada_fig: 3d-disk}
\end{figure}

\section{Gas Dynamics in the Inner 100 pc}
\label{wada_sec: 3}
In order to understand the fueling process from $R\approx 100$ pc to the 
circumnuclear region, we have to understand, at least, the three-dimensional 
structure of the ISM around the central massive black hole at sub-pc resolution.
Instead of adopting a phenomenological approach for the multi-phase ISM (e.g.,
 Ikeuchi, Habe, \& Tanaka 1984; Combes \& Gerin 1985), Wada \& Norman (2001) 
tried to obtain relevant numerical models of the ISM by solving
time-dependent, non-linear hydrodynamical equations and the Poisson equation 
that govern the dynamics and structure of the ISM, using a high-accuracy Euler 
mesh code. They used the advection-upstream splitting method (AUSM; Liou \& 
Steffen 1993), and they achieved third-order spatial accuracy with MUSCL.
AUSM is an improvement of the flux-vector splitting scheme, where the 
advection and pressure terms are separately split at a cell surface. 
(See details in Wada \& Norman 2001.)

Wada \& Norman solve the mass, momentum, and energy equations and the Poisson 
equation numerically in three dimensions to simulate the evolution of a 
rotating gas disk in a fixed, spherical gravitational potential.  The potential
of the stars, dark matter, and a supermassive black hole are assumed to be 
time independent (no feedback from the gas is considered).
The main heating source is supernova explosions. 
Instead of assuming a ``heating efficiency'' to evaluate the dynamical
effect of the blast waves on the ISM, we explicitly follow the evolution of 
the blast waves in the inhomogeneous ISM with sub-pc spatial resolution. 
Radiative cooling is considered not only for the hot gas, but also for 
gas below $10^4$ K, because the pc-scale fine structure of the ISM 
is mainly determined by cold ($T_g < 100$ K), dense media.
Such fine structure is especially important for the fueling processes in
the central 100 pc region.  We assume a cooling function with solar metallicity 
for the temperature range between 20 K and  $10^8$ K.

Figure \ref{wada_fig: 3d-disk} shows the quasi-stable density field of the 
three-dimensional disk model in a central 256 pc $\times$ 256 pc region 
without a central massive black hole. The plot shows the volume-rendering 
representation of density, and the greyscale represents relative opacity. 
As in the two-dimensional models (e.g., Fig. 12 in Wada \& Norman 2001),
the disk shows a tangled network of many filaments and clumps.
Those filaments are formed mainly through tidal interactions between dense 
clumps.  The gas clumps formed by gravitational instability are not
rigid bullets; hence, close encounters between them cause tidal tails, and 
the tails are stretched due to the galactic rotation and local
shear motion. The clumps and filaments collide with each other, and 
this causes the complicated networks. Moreover,  supernova explosions are 
assumed to occur in the model, and their blast waves blow the gas up from the 
disk plane, enhancing the inhomogeneity.

\begin{figure}
\centering
\plotone{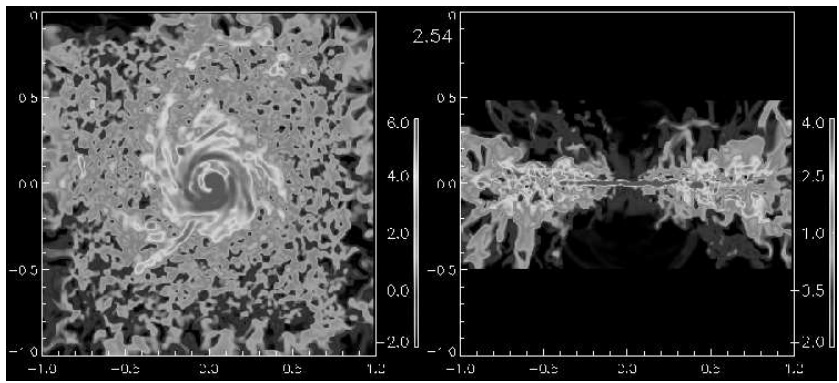}
\caption{Cross sections of density distribution of the gas disk around a 
central massive black hole.  The boxes are 64 pc across. The greyscale 
represents log-scaled density ($M_\odot$ pc$^{-3}$).}
\label{wada_fig: t2}
\end{figure}

For the gas around the AGN, the gravitational potential exerted from
the central massive black hole gives the structure seen in Figure 
\ref{wada_fig: t2}, which displays cross sections of the torus.  The disk has 
a complicated internal structure.  In this model, an average supernova rate is 
assumed to be about 1 supernova per year.  This corresponds to a star 
formation rate of $\sim$100 $M_\odot$ yr$^{-1}$, which would be too high for
typical Seyfert 2s with starbursts (Cid Fernandes et al. 2001; Heckman 2003). 
Note, however, that the scale height of the disk is roughly proportional to 
(SFR)$^{1/2} r^{1.5}$, as discussed later; therefore, the geometry seen in 
Figure \ref{wada_fig: t2} is almost the same for a disk with 2 times larger 
radius and 1/10 of the star formation rate.

\begin{figure}
\centering
\plotone{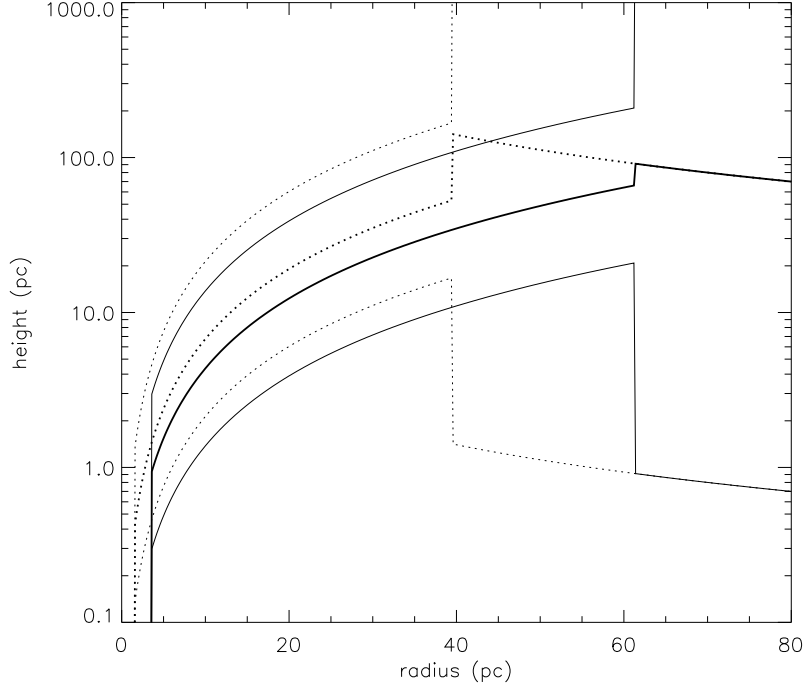}
\caption{Scale height of the disk around a supermassive black hole for three
star formation rates (1, 10, 100 $M_\odot$ yr$^{-1}$) and two black hole mass
(solid lines: $M_\bullet = 1.2 \times 10^8\, M_\odot$; dotted lines :
$M_\bullet =  5 \times 10^7\, M_\odot$). The sharp transitions at $R =$ 40 and
60 pc represent the boundaries of Domain II and Domain III. These transitions
are expected to be smoother for a realistic mass distribution.}
\label{wada_fig: torus}
\end{figure}

Time evolution of the ``torus'' shows that 
the internal motion is not steady, but 
the global concave geometry is supported by internal turbulence
caused by supernova explosions.
In Wada \& Norman (2002), this is explained using a simple analytic 
argument, in which the scale height of the thick disk is
determined by the balance between the turbulent energy dissipation and
the energy feedback from the supernova explosions under the effect of the 
gravitational potential of the supermassive black hole and the stellar disk.
Assuming hydrostatic equilibrium in the vertical direction, we find that the 
radial dependence of the scale height, $h(r)$, is proportional to $r^{3/2}$ in 
the region where the black hole potential dominates ($ r < r_0$),
while it changes as $r^{-1}$ for the outer region ($ r > r_0$). Namely,
\bea
 h_{1}(r) &=& h_{0,1}(r_0) (r/r_0)^{3/2}, \\ 
 h_{2}(r) &=& h_{0,2}(r_0) (r/r_0)^{-1}
\eea
where $r_0 \approx 60 (M_\bullet/10^8\, M_\odot)^{1/2}$ pc 
for a stellar surface density of $10^4\, M_\odot$ pc$^{-2}$, 
and
\bea
h_{0,1}(r_0) &\approx&  35 \, {\rm SFR}_1^{1/2} r_{6}^{3/2} M_{g,8}^{-1/2} 
\,\,\, {\rm pc} 
\eea
and
\bea
h_{0,2}(r_0) &\approx&  1\, {\rm SFR_1}^2 r_{6}^{3} M_{g,8}^{-1} 
\,\,\, {\rm pc}, 
\eea
where the total gas mass $M_{g,8} \equiv M_g/10^8\, M_\odot$, the star 
formation rate ${\rm SFR}_1 \equiv\, 1 M_\odot$ yr$^{-1}$, and 
$r_{6} \equiv r/60$ pc.  The scale heights, $h(r)$, for three star formation 
rates and for two black hole masses are plotted in 
Figure \ref{wada_fig: torus}.  The solutions have three domains: (I) a stable 
disk region ($r < 2-5$ pc), (II) a flared disk
region [$h(r) \propto r^{3/2}, \; 5 < r < 40-60$ pc], and (III) a region where
$h(r) \propto r^{-1}$.
For a less massive central black hole and larger star formation rate,
the disks become thicker. Domain III is more sensitive to the energy
input than Domain II; 
therefore, the scale height of the torus is larger than  1 kpc
for ${\rm SFR}_1$ = 100, which means a ``galactic-wind-like'' solution.
In Domain I, the disk does not fragment to clumps because of the strong shear;
thus, no star formation is expected and the disk should be very thin.

\begin{figure}
\centering
\plotone{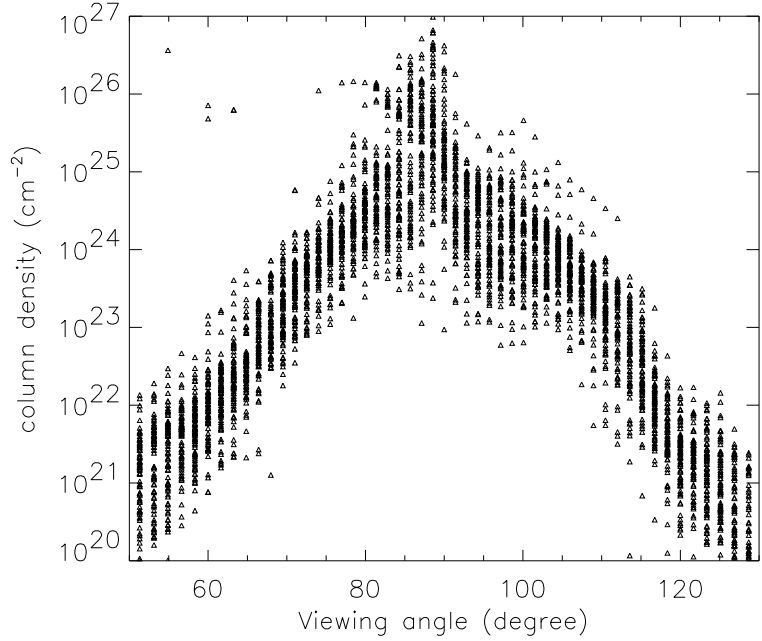}
\caption{Column density distribution as a function of the viewing angle
for the inhomogeneous gas around a supermassive black hole (from Wada 
\& Norman 2002).}
\label{wada_fig: cdensity}
\end{figure}

The column density toward the nucleus as a function of the viewing angle is
plotted in  Figure \ref{wada_fig: cdensity}.  A viewing angle of 
90$^\circ$ is edge-on.  It shows that the viewing angle should be less than 
about $\pm$40$^\circ$ from edge-on to have a large  column density 
($> 10^{23}$ cm$^{-2}$), which is suggested in some Seyfert 2s with nuclear 
starbursts (Levenson, Weaver, \& Heckman 2001). However, one should note here 
that since the internal structure of the torus is very inhomogeneous, the 
column density for the torus is not a simple function of the viewing angle.
There is a large dispersion, $\sim$2 orders of magnitude, in the 
column density for a given viewing angle.

\begin{figure}
\centering
\plotone{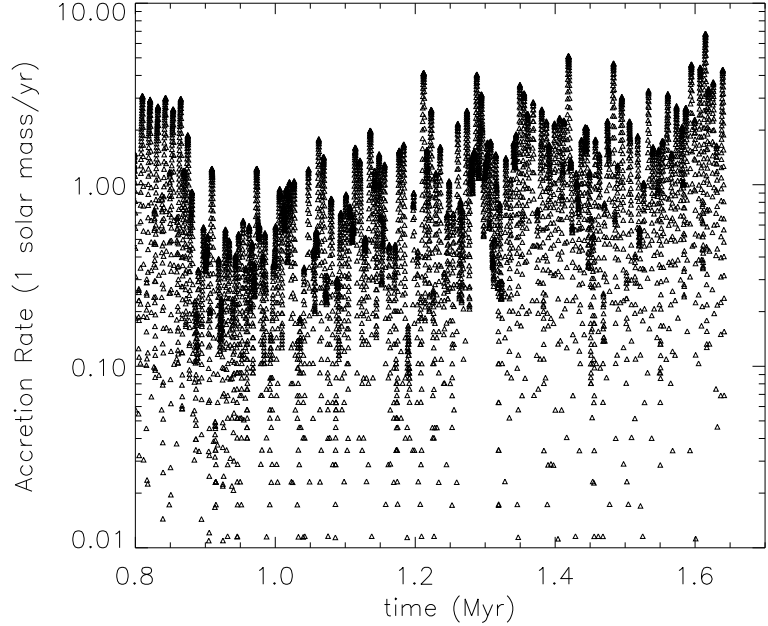}
\caption{Mass accretion rate for $R < 1$ pc from the model shown in 
\S~\ref{wada_fig: t2}.}
\label{wada_fig: mdot}
\end{figure}

The average mass accretion rate for the $R< 1$ pc region in the above-mentioned model
is 0.3 $M_\odot$ yr$^{-1}$, 
and it is about twice as large as that for the model without energy feedback.
Yamada (1994) suggests a positive correlation between X-ray and CO luminosity in
Seyfert galaxies and quasars. If the X-ray and CO luminosity 
correlate with 
the mass accretion rate and star formation rate, respectively, 
the model is qualitatively consistent with the observations.
In Figure \ref{wada_fig: mdot}, 
the gas accretion rate to the nucleus is shown.
An important point here is that the mass accretion to the nuclear region 
should not be a steady flow.
As seen in the plot,
the accretion rate is highly time dependent, with fluctuations over
3 orders of magnitude between 10 and 0.01 $M_\odot$ yr$^{-1}$.
The time scale of the fluctuation is $\sim 10^4-10^5$ yr.
This is a consequence of the ISM in the inner 100 pc being
inhomogeneous. The mass accretion is caused by the kinematic viscosity of
the turbulent velocity field, which is maintained by gravitational
instability and galactic rotation (the same mechanism 
mentioned in \S~\ref{wada_sec: 2.2}); 
the energy feedback from the supernovae enhances the viscosity.
The non-steady inflow for a pc-region would 
be an important feature of accretion 
around a supermassive black hole.

\begin{figure}
\centering
\plotone{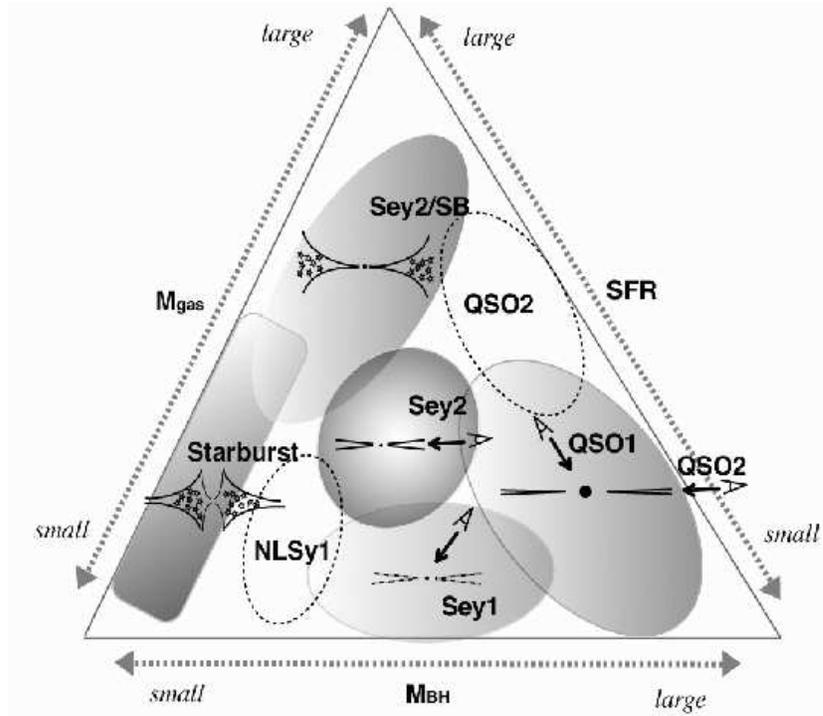}
\caption{Segregation of various types of AGNs from the point of
view of the circumnuclear gas disk.}
\label{wada_fig: agn}
\end{figure}

Suppose that AGNs generally have circumnuclear gas disks with sizes of several 
tens pc and that the global structure of the disk is determined by the 
above-mentioned mechanism.  Various types of the AGNs then could be 
schematically segregated on a plot with three axes, as shown in 
Figure \ref{wada_fig: agn}.  The axes are the total gas mass (or surface 
density of the gas), the black hole mass, and the star formation rate.  Type 2 
Seyferts with starbursts would be gas-rich and their star formation rate is 
high, but the black hole mass would be relatively small.  Therefore, the scale 
height of the disk is large, as shown in Figure \ref{wada_fig: t2}.
On the other hand, the gas disks of Seyfert 1s would be thinner than those of 
Seyfert 2 with starbursts because of a lower star formation rate and/or small 
gas mass. Narrow-line Seyfert 1s might have relatively small black holes.
Quasars would have more massive central black holes 
and a small star formation rate, as a result of which
their circumnuclear disks would be very thin, even if they are gas rich.
Hence, most quasars are observed as type 1, and type 2 quasars 
would be observed by chance, only when the gas disks are edge-on.
However, there might be a counterpart of type 2 Seyferts/starbursts in
the quasar family, in which the nucleus is obscured by the inhomogeneous, dusty
thick disk with active star formation.  Such obscured quasars 
might be the sources of the X-ray background radiation (Fabian 2003).

Finally, I should mention an important physical process that has not 
yet been taken into account for the gas dynamics in the central 100 pc region.
Ohsuga \& Umemura (2001) explored the formation of 
dusty gas walls induced by a circumnuclear starburst around an AGN. 
They found that the radiation force of the circumnuclear 
starburst works to stabilize optically thick walls surrounding the nucleus.
It would be interesting to study the effect of the radiation pressure 
on the dust in the inhomogeneous, turbulent media found in the simulations 
discussed above. Another interesting feature to study are the effects of the
ionizing radiation from the AGN and the starburst regions.
So far, the UV radiation field has been assumed to be uniform, but
apparently this is incorrect in the inhomogeneous ISM around an 
AGN. Solving the radiation field correctly is necessary to compare
models with the abundant available observational information on 
the ionized gas in AGNs. 
Another observational quantity that could be derived from the 
simulations is the molecular line intensity. Work is now in progress 
to incorporate three-dimensional non-LTE radiative transfer for various 
molecular lines for the molecular tori.

\section{Summary: Do We Need Triggers for Nuclear Activity?}
\label{wada_sec: 4}
It has been argued that the fueling problem is essentially 
equivalent to the questions of how we can remove the large angular momentum
of the gas in a galactic disk and how to bring the gas into the small-scale 
region ($R \approx 1$ pc) during the lifetime of an AGN. Non-axisymmetric 
perturbation of the gravitational potential, such as that due to
stellar bars and companions, has been considered as the most plausible 
mechanism. However, this conventional picture should be reconsidered.

The majority of recent observations have not shown clear excess of bars or 
companions in galaxies with AGNs (e.g., Ho, Filippenko, \& Sargent 1997, 2003; 
Mulchaey \& Regan 1997; Corbin 2000; Schmitt 2001). Moreover, recent 
observations of the circumnuclear regions ($ R < 100 $ pc) by the {\it Hubble 
Space Telescope}\ suggest that no significant differences  are found 
in the structure of the nuclear dust lanes between active and inactive 
galaxies (Martini et al. 2003a,b).  Some research groups claim, on the 
contrary, that bars are more abundant in Seyfert hosts than non-Seyfert 
galaxies (Knapen, Shlosman, \& Peletier 2000; Laine et al. 2002).  Although 
this is controversial, one should note that the correlation between bars and 
AGNs is {\it weak} in any results. For example, Laine et al. (2002) found that 
73\% (41 out of 56) of Seyfert hosts are barred, while 50\% (28 out of 56) are 
barred in the control sample. In their small sample, 
the difference between 73\% and 50\% is statistically subtle. 
The result also shows that at least 50\% of non-barred galaxies host AGNs,
and a natural interpretation of this seems to be that AGNs are independent of 
bars. There is another example: Sakamoto et al. (1999) found that the 
concentration factors of the CO, $t_{\rm con} \equiv \Sigma_{\rm gas} 
(R< 500 {\rm pc})/\Sigma_{\rm gas} (R < R_{25})$, where $\Sigma_{\rm gas}$ is 
the average surface density of the molecular gas, for SB$+$SAB galaxies are 
distributed between 20 and 300, while they are between 10 and 70 for 
non-barred galaxies (NGC 4414 has an exceptionally low $t_{\rm con} = 0.9$, 
because CO forms a ring in the galaxy). On average, the degree of gas 
concentration in the central kpc is higher in barred systems than in unbarred
galaxies. One should note again, however,  that {\it more than half} of barred 
and non-barred galaxies have the same range of the concentration factor, 
$t_{\rm con} \approx 20-70$.  Of course, these samples are still too small to 
produce a statistically reliable conclusion, but this can be interpreted as 
indicating that bars have only a weak effect in 
the concentration of the molecular gas.

Suppose the {\it weak} correlation between large-scale perturbation and
nuclear activity is true. How can we explain this theoretically?
One plausible interpretation is that 
bars or companions are one necessary condition for nuclear activity, but that
other conditions related to the pattern speed or strength of the bars, 
gas mass in a certain radius, 
or secondary bars (Maciejewski \& Sparke 2000) need to be
satisfied at the same time in order 
to trigger gas accretion onto the supermassive black hole.
For example, the fraction of the gas mass relative to the dynamical mass 
may need to exceed $\sim$0.1 for fueling through
the collapse of the ILR ring (Wada \& Habe 1992). 
The fueling process also depends on the pattern speed of the bar and/or 
the central rotation curves. Since the ILR ring evolved from leading spirals 
near the IILR, the ring is not formed either for bars with too fast pattern 
speeds [i.e., $\Omega_p > {\rm max} (\Omega - \kappa/2)$]
or for a stellar potential with a central cusp.
As mentioned in \S~\ref{wada_sec: 2.1},
recent studies suggest that most spiral galaxies have a central mass 
concentration rather than a flat core in the central kpc region. 
If this is the case, an OILR or nILR is expected.
Therefore, the trailing spiral shocks or offset ridges are 
the probable structures expected in the central parts of
barred galaxies. In this case, gas fueling for the nucleus 
can be stopped at a certain radius inside the OILR or nILR, and its
radius depends on the amount of angular momentum and energy 
loss at the trailing shocks\footnote{Maciejewski et al. (2002) showed that
for high sound speed a spiral shock propagates to the center, which may be
responsible for the fueling.}. Since gravitational instability of the ring
is a trigger for further inflow, in this case the gas mass, the strength of 
the bar, and the pattern speed of the bar are the additional factors needed 
for the onset of fueling.
Fukuda et al. (2000) also suggested that energy feedback from 
supernovae in the ring enhances gas accretion to the nucleus.
As mentioned in \S~\ref{wada_sec: 3}, energy feedback from star formation
in the nuclear region is another important factor in determining the 
mass accretion rate.

Besides the gravitational torque from a non-axisymmetric potential, 
dynamical friction between the gas clumps and the stellar system (Shlosman \&
Noguchi 1993), the viscosity due to clump-clump collisions (Ozernoy, Fridman, 
\& Biermann 1998; Begelman, Frank, \& Shlosman 1989), gravity-driven 
turbulence (Lynden-Bell \& Pringle 1974; Paczy\'nski 1978; Wada et al. 2002)
and supernova-driven turbulence (von Linden et al. 1993) are important fueling 
mechanisms  that do not involve bars (see Fig. \ref{wada_fig: diagram}).
In addition to these, radiative avalanche (Umemura, Fukue, \& Mineshige 1998; 
Umemura 2003; Kawakatsu, Umemura, \& Mori  2003), magneto-hydrodynamical 
turbulence (Balbus \& Hawley 1991; for a recent three-dimensional global 
magneto-hydrodynamical simulation of an accretion disk, see Machida, Hayashi, 
\& Matsumoto 2000 and Machida \& Matsumoto 2003), and spiral density waves 
(Goldreich \& Lynden-Bell 1965a,b; Lynden-Bell \& Kalnajis 1972) may also be 
important.  In these mechanisms, the onset of mass accretion is controlled by
many factors, such as the structure of the ISM, stellar mass density, 
star formation rate, the initial mass function, the dust-to-gas ratio, 
strength of the magnetic field, ionization fraction, the total gas mass, etc.
In light of these various factors for the onset of gas accretion, 
a stellar bar may not even be a necessary condition for gas fueling.

Finally, I would like to emphasize two important {\it scales} concerning
the fueling problem: the spatial scale and the time scale.
AGNs are powered by mass accretion onto supermassive 
black holes, on scales of $R \approx 10^{-5}$ pc. 
In order to explain the enormous luminosity of AGNs,
we should treat the accretion phenomena on small scales.
However, most studies of gas fueling triggered by
galactic-scale phenomena, such as bars, mergers, and interactions,
are focused on gas dynamics in regions 100--1000 pc from the galactic center. 
Apparently, 
mass accretion to such a region does not mean that the accumulated gas 
can fall all the way to the accretion disk.
In \S~\ref{wada_sec: 3}, I have shown some new results
on the gas dynamics between these two regimes, on scales 1--100 pc.
We do not yet have enough observational information on this scale.
For example, the structure and dynamics of the molecular gas in
the galactic central region have been explored by 
radio interferometers with $\sim 1^{\prime\prime}$ resolution, but this is not
fine enough to resolve the inhomogeneous 
structure of the ISM, expected in the numerical simulations, 
even for nearby galaxies  (Wada \& Koda 2001). 
The next generation radio interferometer 
ALMA (Atacama Large Millimeter/submillimeter Array) will be the instrument for 
exploring the ``missing link'' in the fueling problem. 
The 0\farcs01 resolution achievable by ALMA 
can reveal the sub-pc structure of the
molecular gas in the central 100 pc of nearby galaxies 
(e.g. those in the Virgo cluster). 

Time variability is another important feature of AGNs.  The average lifetime 
of AGN activity is $\sim 10^7$ yr (Martini 2003). This does not mean, however, 
that the mass accretion (e.g., $1\, M_\odot$ yr$^{-1}$), is constant during 
this lifetime.
As mentioned in \S~\ref{wada_sec: 2.2},
we expect that the turbulence is self-regulated in a dense gas disk (Wada et al. 2002),  
and the turbulence driven by self-gravity in an inhomogeneous ISM causes 
stochastic mass accretion. 
This would be very important as an outer boundary condition for
the accretion disk around the supermassive black hole. 
Our results suggest that the time scale of non-steady mass accretion 
is $\sim 10^{4-5}$ yr, which would be much shorter than the lifetime
of AGNs. If this is the case, any galaxies with a massive gas core and
supermassive black hole can be active, and they might
be recognized as luminous AGNs only for that short period.
 Observations with ALMA are again essential 
to reveal the kinematics of the ISM in the central 100 pc, and to couple this 
information to the evolution of AGNs.

\subsection*{Acknowledgment}
I would like to thank Colin A. Norman, Asao Habe, and Jin Koda for our 
collaboration.  I am also grateful to Luis Ho for organizing this fruitful 
conference, and to Witold Maciejewski for his valuable comments on the draft. 
Our numerical simulations were performed on the supercomputer system in the
Astronomical Data Analysis Center, National Astronomical Observatory of Japan.

\end{document}